\documentclass[prd,nofootinbib,superscriptaddress,twocolumn]{revtex4}

\usepackage{graphicx}
\usepackage[breaklinks=true,colorlinks=true,linkcolor=blue,urlcolor=blue,
citecolor=blue]{hyperref}

\usepackage[mathscr]{eucal}
\usepackage[latin1]{inputenc}
\usepackage{amsmath}
\usepackage{amsfonts}
\usepackage{amssymb}
\usepackage{pictex}

\def\db{\delta_b}
\def\dc{\delta_c}

\def\be{\nopagebreak[3]\begin{equation}}
\def\ee{\end{equation}}
\def\ba{\nopagebreak[3]\begin{eqnarray}}
\def\ea{\end{eqnarray}}

\newcommand{\teta}{\rlap{\lower2ex\hbox{$\,\tilde{}$}}\eta{}}

\def\lp{{\ell}_{\rm Pl}}

\usepackage{enumerate}

\begin{document}

\title{Von-Neumann Stability and Singularity Resolution in Loop Quantized \\ 
Schwarzschild Black Hole}
\author{Alec Yonika}
\affiliation{Department of Physics \& Center for Scientific Computing and 
Visualization Research,
      University of Massachusetts Dartmouth, North Dartmouth, MA 02747, USA}
\author{Gaurav Khanna}
\affiliation{Department of Physics \& Center for Scientific Computing and 
Visualization Research,
      University of Massachusetts Dartmouth, North Dartmouth, MA 02747, USA}
\author{Parampreet Singh}
\affiliation{Department of Physics and Astronomy, \& Center for Computation and 
Technology,
 Louisiana State University, Baton Rouge, LA 70803, USA}


\begin{abstract}
Though loop quantization of several spacetimes has exhibited existence of a 
bounce via an explicit evolution of states using numerical simulations, the 
question about the way central singularity is resolved in the black hole interior has remained open. 
The quantum Hamiltonian constraint in loop quantization turns out to be a finite difference equation whose stability 
is important to understand to gain insights on the viability of the underlying quantization and  resulting physical implications. We 
take first steps towards addressing these issues 
for a loop quantization of the Schwarzschild interior recently given by Corichi and Singh. Von-Neumann stability 
analysis is performed using separability of solutions as well as a full two 
dimensional quantum difference equation. This results in a stability condition for black holes which have a very large mass compared 
to the Planck mass. For black holes of smaller masses evidence of numerical instability is found. 
In addition, stability analysis for macroscopic black holes leads to a constraint on the choice of the allowed 
states in numerical evolution. States which are not sharply peaked in accordance with this constraint result in instabilities. With the caveat of using kinematical 
norm, sharply peaked Gaussian states are evolved using the quantum difference equation and 
singularity resolution is obtained. A bounce is found for one of the triad 
variables, but for the other triad variable singularity resolution amounts to a 
non-singular passage through the zero volume. States are found to be peaked at 
the classical trajectory for a long time before and after the singularity resolution, 
and retain their semi-classical character across the zero volume. Our main result is that quantum bounce 
occurs in loop quantized Schwarzschild interior at least for macroscopic black holes. Instability of small black holes which can be a 
result of using kinematical norm nevertheless signifies the need of further understanding of the 
viability of the considered quantization and its physical Hilbert space.

\end{abstract}

\maketitle

\section{Introduction}
Existence of physical singularities in classical gravity is often tied to the 
underlying continuum differentiable spacetime manifold. As vanishingly small 
scales are probed, spacetime curvature grows unboundedly eventually resulting in 
a divergence at the classical singularity where the evolution breaks down. The 
hope in quantum theories of gravity is that incorporating the quantum discreteness 
of spacetime results in a resolution of singularities. In the last decade, 
this hope has been realized for the many spacetimes quantized using techniques of 
loop quantum gravity~\cite{mb-livrev,as-status}. Here, thanks to the quantization 
procedure, the classical differential geometry is replaced by a discrete quantum  
geometry where geometrical operators have discrete eigenvalues with a non-zero 
minimum. The quantum discreteness becomes significant only in the Planck regime, 
and results in a classical continuum spacetime at larger scales. In all the 
spacetimes where a rigorous loop quantization has been performed, the above 
picture results in the resolution of the physical singularity at the Planck 
curvature scale. An agreement with general relativity (GR) is found at the 
scales when curvature becomes much smaller. The success is notable for cosmological 
models where the big bang is replaced by a big bounce~\cite{aps3}. States 
peaked at classical trajectories in a large expanding macroscopic universe,  
when evolved backward towards the big bang using the discrete quantum evolution 
equation, follow the classical trajectory for a long time, and bounce in the Planck 
regime to a contracting branch. Quantum evolution is {\it{stable}} and {\it{non-singular}}, 
and the results of existence of a bounce hold even if the states have very wide spreads and 
large fluctuations~\cite{wide}. 
 
In the isotropic cosmological models, singularity resolution in loop quantum cosmology (LQC) 
has been established using analytical \cite{slqc,craig} and numerical investigations (for a 
review see Refs. \cite{BriCarKha2012,Sin2012}). There also have been numerical studies on 
Bianchi-I spacetimes~\cite{bianchi-madrid,bianchi-lsu}, which confirm the resolution of singularities. 
But, such numerical investigations are 
so far absent for black hole spacetimes. And, unlike the isotropic model, it is quite difficult to solve 
anisotropic and black hole spacetimes analytically, except in an effective spacetime description where strong hints are present   for  a 
generic resolution of strong curvature singularities \cite{generic}.  Especially in the black hole case, given the technical difficulties, 
none of the loop quantization attempts has so far addressed the construction of the physical Hilbert space. Hence, even numerical 
studies need to be performed under certain assumptions such as the unavailability of the physical inner product. These numerical 
studies are in any case quite non-trivial. This is because the loop quantization of black hole 
spacetimes results in a quantum Hamiltonian constraint, which is a coupled 
quantum difference equation in two triad variables. The discreteness in both the variables is fixed by the 
underlying quantum geometry and a priori there is little guarantee that the evolution is stable. Here, we recall that the loop 
quantization of the Schwarzschild interior is performed using a Kantowski-Sachs 
vacuum spacetime with a phase space expressed in terms of holonomies of 
Ashtekar-Barbero connection components $b$ and $c$, and the two conjugate triad 
variables $p_b$ and $p_c$. Here $p_c$ is proportional to $g_{\Omega \Omega}$, and $p_b^2/p_c$ is proportional to $g_{xx}$ components of the metric. (For more details, see Sec. II).  Due to the associated quantization ambiguities, 
different quantization prescriptions for the Schwarzschild interior 
exist~\cite{ab-bh,modesto,CarKha2006,bv,gp1,bck,SabKha2007,gp2,CorSin2016}. These 
amount to variations in the way minimum area loops are constructed with different 
triad dependencies. Of these prescriptions, a recently proposal  by Corichi 
and Singh is notable \cite{CorSin2016}. Unlike other attempts, it results in a consistent 
infra-red limit as GR and is free from fiducial structures used in the quantization 
procedure. Quantum Hamiltonian constraint of this model, here after 
referred to as the CS model, is known to be non-singular in the sense that one can 
analytically show that initial data can be propagated through the classical 
singularity at $p_c = 0$,  at least at the kinematical level. But as with any 
other loop quantization of the Schwarzschild interior, explicit singularity 
resolution via evolution of quantum states has so far not been shown. 
 
To understand the consistency of the quantum Hamiltonian constraint and the resulting physics, 
one needs to ensure that the quantum difference equation is von-Neumann stable. Only then the resulting 
quantization has a well defined classical limit. This is a non-trivial problem in 
loop quantum gravity. To address this problem, one has to overcome some challenges which are not found in  classical 
numerical relativity. In the latter, given the underlying continuum spacetime manifold of GR, one has freedom to appropriately 
choose discreteness in space and time grids to get a stable evolution. Due to the underlying quantum geometry, this freedom is absent in 
loop quantum gravity. Since the underlying discreteness of `space' degrees of 
freedom is fixed by quantum geometry, there is no guarantee whether the evolution 
is intrinsically stable or unstable. This is required to be verified explicitly 
by understanding the growth of the solutions in a `time' degree of freedom on 
the stencil provided by the finite difference quantum Hamiltonian constraint. 
Examples of quantum constraints yielding stable and unstable evolution exist in 
LQC~\cite{bck,BriCarKha2012,Sin2012}. It turns out that the quantizations where 
von-Neumann stability is problematic, there are often other problems with the 
quantization prescription which independently reveal them to be inconsistent~\cite{prob}. 
On the other hand, when the quantum constraint turns out to be stable, 
conditions for existence of classical solutions are fulfilled~\cite{preclassicality}, 
ensuring the agreement of infra-red limit obtained at large scales  
with the classical theory. However, stability of the quantum difference equation, 
via the Courant-Friedrichs-Lewy condition, imposes further requirements on the 
choice of numerical discreteness used for any continuous variable playing the role of 
intrinsic time variable. An example is in the case of a massless scalar field 
coupled to gravity in isotropic LQC where new algorithms, or high performance 
computing becomes necessary to overcome associated computational challenges~\cite{chimera}. 
Fortunately, these problems do not arise for the Schwarzschild interior where the 
role of the clock can be played by one of the directional triad variables. But, an 
additional difficulty in comparison to isotropic LQC is that the clock as well as 
the other directional triad being measured are both discretized due to the underlying 
quantum geometry. This leaves no wiggle room for the quantum evolution to be stable.

The goal of our analysis is two fold. First to investigate the von-Neumann 
stability  in the loop quantization of the Schwarzschild interior in the 
CS model, and second to gain insights on the dynamical resolution of singularity 
by evolving states peaked at late times towards the classical central singularity. 
The latter analysis is based on the caveat that as in the case of all other loop 
quantizations of the Schwarzschild spacetime, an inner product or the physical Hilbert 
space is not known. Rather we use a kinematical $L^2$ norm to understand the behavior 
of the expectation values of one of the triads $p_b$ with respect to the 
other triad $p_c$. Due to this reason, this part of our analysis can be 
considered as a first step towards investigating singularity resolution using numerical 
simulations for the loop quantized black holes. Let us note here that it is not for the first time that a kinematical norm is being used in LQC. 
Early works in LQC on isotropic cosmological models used the kinematical norm and captured semi-classical regime with a reasonable success 
in the sense that 
there was agreement with the effective dynamics \cite{coordinate,emergence}. More recently, in isotropic models normalization of the wavefunction valid only for the massless scalar field case \cite{aps3,slqc}, has been used to capture the behavior 
for models with positive and negative potentials yielding a good agreement with the effective dynamics \cite{aag,cyclic}. The latter results show that even without using the true physical inner product, crucial features of singularity resolution can be 
deciphered. Our use of kinematical norm is in the same spirit as above works. The hope is that it might provide 
some useful clues on singularity resolution in the black hole interior.

The main results from our analysis are the following. We find that in the 
approximation of studying von-Neumann stability on values of triads much larger than 
the Planck values the quantum difference equation is stable in the limit where one 
of the two quantum discreteness parameter vanishes or equivalently the mass is 
extremely large compared to the Planck mass. The limit is consistent with the 
underlying approximation because it is only for the large mass black holes that the 
triads can take large values. However, exponential growth is numerically found to be present 
for masses which are not large. This implies that black holes with smaller masses are unstable in the CS quantization. This instability can have many causes. It might be caused by  
the lack of physical Hilbert space in our analysis and not knowing the spectrum of the quantum Hamiltonian constraint operator which does not allow correctly choosing the eigenfunctions for allowed states in numerical simulations. 
As we discuss later in Sec IV, such an example of instability \cite{green}, and its resolution \cite{apsv} already exists in LQC. Given the main caveat of our analysis, the lack of the physical Hilbert space as in other loop quantization of black hole spacetimes, it is not possible to address this issue in this manuscript. 
However, if this instability is caused independent of the above issue,  then it points to 
the lack of viability of CS quantization for the small black holes.  

Interestingly, an additional constraint is found 
on the allowed values of the triads which translates to a restriction on the kind of 
states one can choose in a stable evolution. Using such states with a Gaussian profile 
we find that the expectation values of the triad $p_b$ reach a non-zero minimum value 
in a relational evolution with respect to the (radial) triad $p_c$ which acts as a 
clock. Hence, there is a bounce of the triad $p_b$ with respect to $p_c$. States 
pass through the singularity at $p_c = 0$, from positive values to the negative 
values, in a non-singular way without any breakdown of the quantum evolution. 
States grow in fluctuations near the classical singularity but retain their 
characteristics quite symmetrically across the resolved singularity. At early times,  
beyond the singularity resolution, a typical state regains its sharply peaked nature, 
and is peaked on a classical trajectory as is the state at late times. In fact, 
states remain sharply peaked on the classical trajectory till they reach Planck 
scale where a striking departure from the classical theory occurs due to the 
underlying quantum geometry. 

The manuscript is organized as follows. In Sec. II, we start with the quantum 
difference equation of the CS model and using separation of variables and express 
it as two uncoupled quantum difference equations in two independent triads. 
Using the approximation of large values of triads, in Sec. IIA we analytically 
obtain the limiting behavior of solutions of these difference equations which 
shed light on the pre-classicality of these equations. In particular we find 
that the discreteness parameter associated with the holonomies of the connection 
component conjugate to $p_b$ must vanish, implying that within the numerical 
precision the mass of the black hole must be very large compared to the Planck scale.  
In Sec. IIB, this analysis is repeated using numerical techniques and stability 
is studied yielding the same conclusion. A comparison of analytical and numerical 
methods for separable solutions and their close agreement is also discussed in Sec. IIB. 
Gaining insights from the 1-D behavior of the quantum difference equation for 
the CS model we perform a stability analysis using the full 2-D quantum difference 
equation in Sec. III. The result turns out to be the same along with a restriction  
that $k < 4 n$ where $k$ and $n$ are measures of $p_b$ and $p_c$ respectively. 
Further, in this section we study the behavior of states using $n$ as a clock
and find the results on singularity resolution. We conclude with a summary of 
results and their discussion in Sec. IV.


\section{Separable solutions}

In this section, we study the space of separable solutions of the quantum Hamiltonian constraint resulting from the loop quantization of Schwarzschild 
interior  as presented in Ref.~\cite{CorSin2016}. We first briefly summarize the main steps leading to the quantum constraint. 
As noted before, in the Ashtekar-Barbero phase space, the gravitational phase space variables 
are conjugate pairs $(b,p_b)$ and $(c,p_c)$. The spacetime metric in terms of the triad variables is given by
\begin{equation}
 {\mathrm{d}} s^2 = - N^2 {\mathrm{d}} t^2 + \frac{p_b^2}{|p_c| L_o^2} {\mathrm{d}} x^2 + |p_c| ({\mathrm{d}} \theta^2 + \sin^2 \theta {\mathrm{d}} \phi^2) ~.
\end{equation}
Here $L_o$ is a fiducial length scale in the $x$-direction of the spatial manifold which has topology $\mathbb{R}\times\mathbb{S}^2$. The fiducial length scale is necessary to be introduced to define the symplectic structure. In terms of the conventional form of the Schwarzschild metric, $p_b$ and $p_c$ satisfy:
\begin{equation}
 \frac{p_b^2}{p_c} = \frac{2 m}{t} - 1. ~~~~ |p_c| = t^2 ~,
\end{equation}
where $m = G M$, with $M$ as the ADM mass of the black hole spacetime. In the classical theory, the horizon at $t=2m$ is identified with 
$p_b = 0$ and $p_c = 4 m^2$ (its maximum value). From the classical solutions one finds that at the horizon, $b$ vanishes and $c = \gamma 
L_o/4 m$, where $\gamma \approx 0.2375$ is the Barbero-Immirzi parameter. The classical evolution breaks down at the central singularity 
where both $p_b$ and $p_c$ vanish, and their conjugates $b$ and $c$ diverge. The triad $p_b$ takes its maximum value equal to 
$p_b^{\mathrm{max}} = m$ in the interior spacetime when 
$b$ takes the value equal to $\gamma$ \cite{CorSin2016}.

The classical Hamiltonian constraint is given by 
\begin{equation}
 C_H = - \int\mathrm{d}^3 x \, e^{-1} \varepsilon_{ijk} E^{ai} E^{bj}(\gamma^{-2} F_{ab}^k - \Omega_{ab}^k)
\end{equation}
where $e$ denotes the determinant of the triads $E^{ai}$ whose symmetry reduction yields components $p_b$ and $p_c$, 
$F_{ab}^i$ is the field strength of the Ashtekar-Barbero connection and $\Omega_{ab}^i$ is the curvature corresponding to the spin 
connection. In the loop quantization, the classical Hamiltonian constraint is expressed in terms of the holonomies of the connection 
components $b$ and $c$. The holonomies of $b$ are considered over edges labeled by $\mu$ in angular directions, where as the holonomies of 
$c$ are labeled by $\tau$ in the $x$-direction. 

In the quantum theory, the eigenvalues of triad operators are given by:
\begin{equation}\label{pb_pc_ev}
\hat p_b \, |\mu,\tau\rangle = \frac{\gamma \lp^2}{2} \, \mu \, |\mu,\tau\rangle, ~~ \hat p_c \, |\mu,\tau\rangle = \gamma \lp^2 \, \tau \, |\mu,\tau\rangle ~.
\end{equation}
Loop quantization of the classical Hamiltonian constraint using the holonomies 
of the connection components $b$ and $c$ yields the following quantum difference equation~\cite{CorSin2016}:
\begin{eqnarray}\label{ham}
&& \nonumber (\sqrt{|\tau|} + \sqrt{|\tau + 2 \dc|}) \left(\Psi_{(k + 2) \db, \tau + 2 \dc} - 
\Psi_{(k - 2) \db, \tau + 2 \dc} \right) \\
&& \nonumber \hskip-0.2cm + \tfrac{1}{2} \, (\sqrt{|\tau + \dc|} - \sqrt{|\tau -  \dc|}) \bigg[(k+2)\Psi_{(k + 4) \db, \tau} \\
&& \nonumber \hskip0.4cm ~~~~~ +  (k-2) 
\Psi_{(k - 4) \db, \tau}  - 2 k (1 + 2 \gamma^2 \db^2)  \Psi_{k \db,\tau} \bigg]\\
&& \nonumber \hskip-0.2cm + (\sqrt{|\tau|} + \sqrt{|\tau - 2 \dc|}) \left(\Psi_{(k - 2) \db, \tau - 2 \dc} - 
\Psi_{(k + 2) \db, \tau - 2 \dc} \right) \\
&& = 0 ~.
\end{eqnarray}
Here, for a Schwarzschild black hole interior corresponding to mass $m$,  
\begin{equation}\label{dbdc}
\delta_b = \frac{\sqrt{\Delta}}{2 m}, ~~~ \mathrm{and} ~~~~ \delta_c = \frac{\sqrt{\Delta}}{{L_o}} ~~~ 
\end{equation}
where $\Delta$ denotes the minimum area eigenvalue in loop quantum gravity: 
$\Delta = 4 \sqrt{3} \pi \gamma \lp^2$. Above, $k = \mu/\db$ with $k > 4$ a measure of eigenvalues 
of $\hat p_b$. 

Starting from the quantum Hamiltonian constraint \eqref{ham}, we extract a 
number of important features using separability that appear to hold more generally. 
For the main part of this section, these correspond to the stability conditions, but we 
also find  that the CS model appears to resolve the $\tau = 0$ (classical) singularity 
cleanly. And the evolution passes through it without any sign of a breakdown of the 
quantum dynamics. 

Let us begin with performing a very traditional separation-of-variables solution of the 
CS model under consideration. In particular, for \eqref{ham}, define $\tau = n\delta_c$ 
and then substitute the ansatz $\Psi_{n,k} \rightarrow A_k B_n$. This allows us to perform 
a separation of variables on the two dimensional partial difference equation and cast 
into a system of two simpler sequences: 
\begin{eqnarray}
(k+2)A_{k+4} & + &(k-2)A_{k-4}= \nonumber \\ 
2k(1+2\gamma^2\delta_{b}^2)A_{k} & + & 2\lambda(A_{k+2}-A_{k-2})
\label{Aeqn}
\end{eqnarray}
and
\begin{eqnarray}
(\sqrt{|n|}+\sqrt{|n+2|})B_{n+2}& - &(\sqrt{|n|}+\sqrt{|n-2|})B_{n-2} \nonumber 
\\ 
=-\lambda(\sqrt{|n+1|}& - &\sqrt{|n-1|})B_{n}
\label{Beqn}
\end{eqnarray}
where $\lambda$ is the separation parameter. Note that these equations are 
similar to the ones presented in Refs.~\cite{SabKha2007} and \cite{CarKha2006}.  

\subsection{Large $n,k$ limit}
We begin our study of these sequences by considering them in the large $n,k$ 
limit. This essentially translates to examining the CS model's semi-classical 
behavior. Note that in this limit, we expect the quantum solutions to take a smooth 
form and resemble the solutions from a differential equation. This expectation is 
connected to the notion of {\em preclassicality} that was studied in depth in 
research papers~\cite{preclassicality} over a decade ago, and recently reviewed 
in Ref.~\cite{BriCarKha2012,Sin2012}. 

To derive the limiting behavior associated to the sequences above in the large 
$n,k$ limit, we perform the substitutions $B_{n} \rightarrow B(n)$ and  
$A_{k} \rightarrow A(k)$, then do Taylor series expansions to second-order in 
inverse powers of $n,k$ and finally substitute those expressions into the system 
\eqref{Aeqn} and \eqref{Beqn}. This results in two relatively simple ordinary 
differential equations that can be solved analytically to obtain: 
\begin{equation}
B(n)=Cn^{-\frac{\lambda+2}{8}}
\end{equation}
and
\begin{eqnarray}
A(k)=k^{\frac{\lambda}{4}}J_{\frac{\lambda}{4}}(-\frac{1}{2}i\gamma\delta_{b} 
k)&C_1& \nonumber \\
+ k^{\frac{\lambda}{4}}Y_{\frac{\lambda}{4}}(-\frac{1}{2}i\gamma\delta_{b} 
k)&C_2&
\end{eqnarray}
where $J$ and $Y$ are Bessel functions of the first and second kind. It is 
interesting to note that the presence of the $Y$ function generates unbounded 
exponential growth generically. This manifests itself in the form of a generic 
instability that appears in the $A$ sequence itself (as is discussed in later 
sections) and also the full model as a whole. 

It is also worth noting that if in the $A$ sequence \eqref{Aeqn}, one 
specific term i.e. $\gamma^2 \delta_{b}^2$ could be made arbitrarily small, then 
this growth can be fully controlled. As an extreme form of that, let us 
simply eliminate that term from the \eqref{Aeqn}. Then the resulting equation  
yields: 
\begin{equation}
A(k)=\frac{2k^{\frac{\lambda}{2}}}{\lambda}C_1+C_2
\end{equation}
Thus, controlling this unstable growth simply requires the condition that 
$\delta_b \rightarrow0$. Using \eqref{dbdc}, this translates to the mass of the 
black hole to be effectively extremely large compared to Planck mass.

\subsection{Numerical recursion solutions}
In this subsection, we perform a detailed study of the solutions of the system 
\eqref{Aeqn} and \eqref{Beqn} taking a fully numerical approach. The numerical 
scheme we utilize is a simple recursion, stepping forward in the variables $n,k$ 
starting with some arbitrary initial values. We graph the results in Fig.~\ref{ABsoln}. 

As can be seen immediately, there are serious instabilities in our model for 
the $A$ sequence. As argued before, this can be attributed to the presence of 
that $\gamma\delta_b$ term. In addition, we also note that the solutions exhibit 
strong high-frequency oscillatory behavior that lingers even into the large $k, n$ 
regime. This suggests that these solutions could not possibly be compatible with the 
differential equation solutions presented in the previous section. One approach to 
address this issue is via the notion of {\em preclassicality} as suggested 
before. That involves selecting a careful set of initial values (as opposed to 
arbitrary ones) that result in smooth solutions in the large $k, n$ regime. The 
procedure to make such a good selection of initial values may be found in Ref.~\cite{CarKhaBoj2004} 
using {\em generating function} techniques. However, we take a different approach 
in this work. 

\newpage

\begin{widetext}
\begin{center}
\begin{figure}[htb!]
\includegraphics[width=0.45\columnwidth]{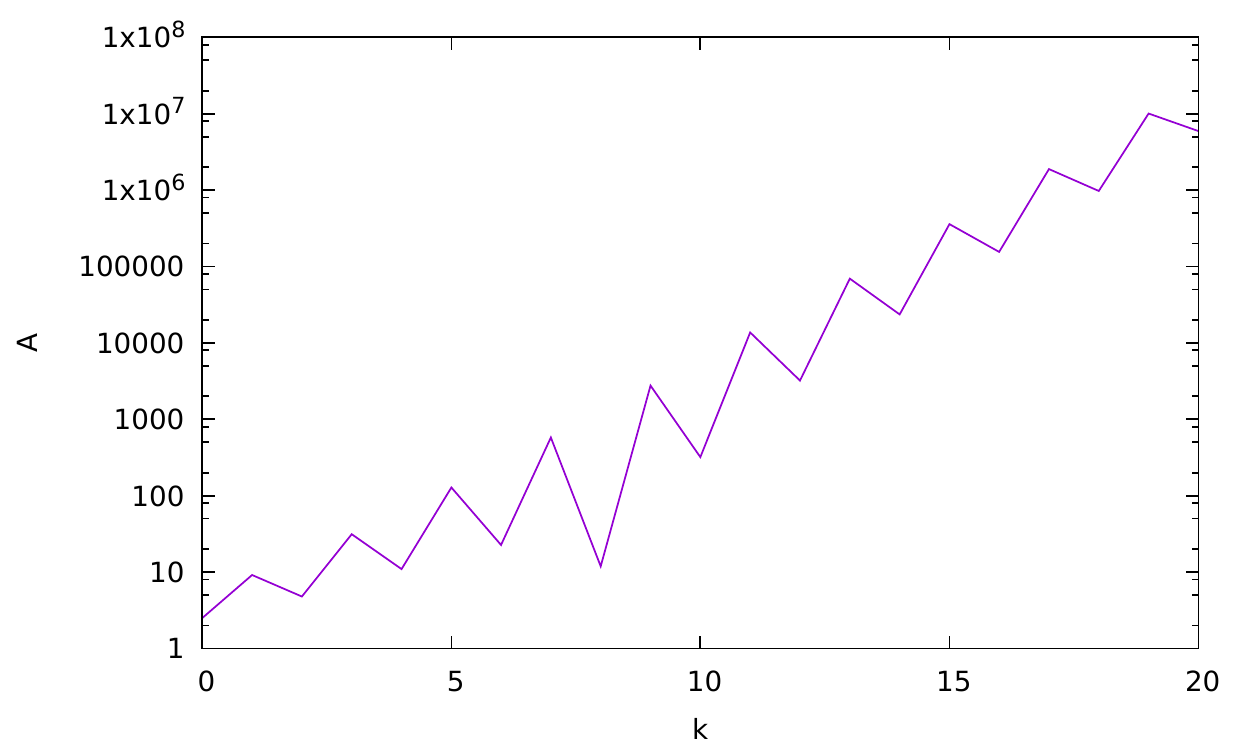}
\includegraphics[width=0.45\columnwidth]{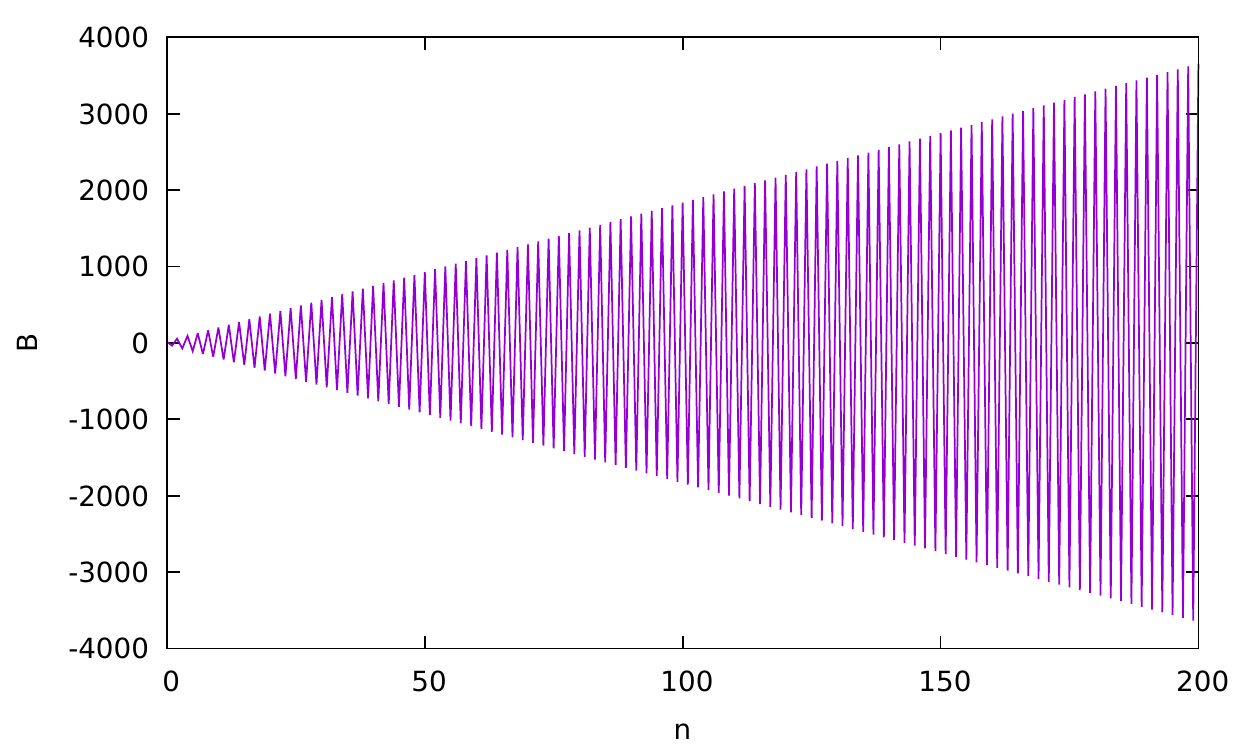}
\caption{Solutions of separated, recursive schemes with arbitrary initial 
conditions.}
\label{ABsoln}
\end{figure}
\end{center}
\end{widetext}

\subsubsection{1D stability analysis of A and B sequences}
Noting that there are instabilities generically in the $A$ sequence, we now 
turn our attention to uncovering the root cause of this undesirable behavior. 
In order to do so we perform  von-Neumann stability analysis which is a standard 
technique used in numerical analysis for finite-difference schemes. Traditionally, 
von-Neumann analysis is done on a function of multiple variables (usually, 
discretized time and space). Since we have two separated equations, 
each of a single variable, we implement a von Neumann stability analysis {\em 
inspired} technique. To do this we introduce an ansatz of the form 
$A_{k} \rightarrow v^{k}$. The resulting expression is then manipulated to obtain 
a polynomial in $v$, of which the roots are found. If the values of the roots are 
less than or equal to one, then the solution is expected to be stable. We enforce 
this condition, and thus obtain a choice of parameters that lead to that condition being satisfied.

The roots of the ``amplitude'' $v$ for the A sequence after performing the 
manipulations mentioned above are:
\begin{equation}
v=(1+2^2\gamma^2\pm 
2\sqrt{\delta_{b}^2\gamma^2+\delta_{b}^4\gamma^4})^{\frac{1}{4}}
\end{equation}
Note that the above expression has been simplified by presenting it in a large 
$k$ limit. Thus, enforcing the condition that $v\leq 1$ results in the constraint: 
\begin{equation}
\delta_b \rightarrow 0. 
\end{equation}
This result should not be a surprise. After all, we argued in the previous 
subsection that this $\gamma\delta_b$ term causes a generic exponential growth via 
the Bessel function of the second kind. Once again, we can arbitrarily reduce 
the rate of the unstable growth by manipulating the value of the $\delta_b$. 

A similar study of the $B$ sequence does not yield any constraints i.e. the 
solutions therein are generically stable.

\subsubsection{Comparison of numerical and analytical solutions}
Given an understanding of the root cause of the instabilities in the $A$ 
sequence, in this subsection, we simply remove that term and compare the results 
with the differential equation solutions presented in the previous sections. This 
allows us to directly compare the results of the quantum dynamics with the semi-classical 
case and also study the regimes where they differ significantly. 

\begin{figure}[htb!]
\centering
\includegraphics[width=\columnwidth]{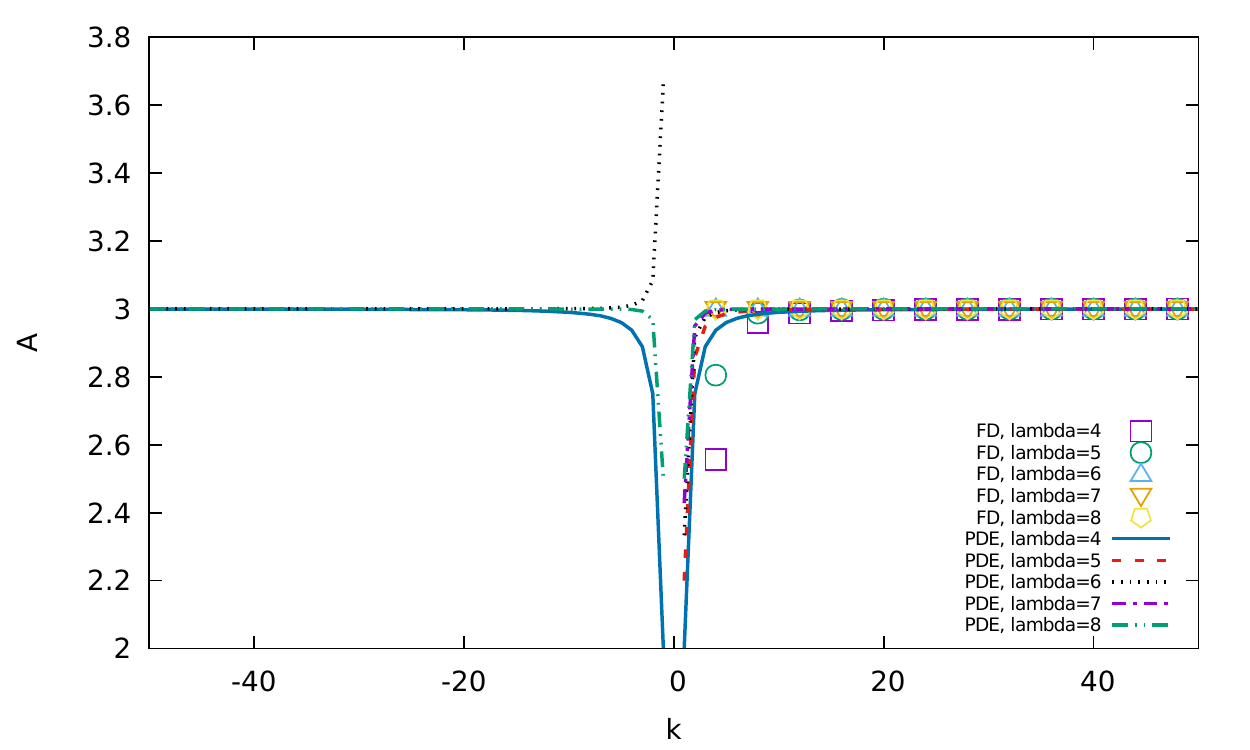}
\caption{Comparison of the numerical recursive solution with the smooth ODE 
solution of the $A$ sequence, with $\lambda = 4,5,6,7,8$. 
Lines are ODE solutions, dots are numerical solutions.}
\label{Acomp}
\end{figure}

\begin{figure}[htb!]
\centering
\includegraphics[width=\columnwidth]{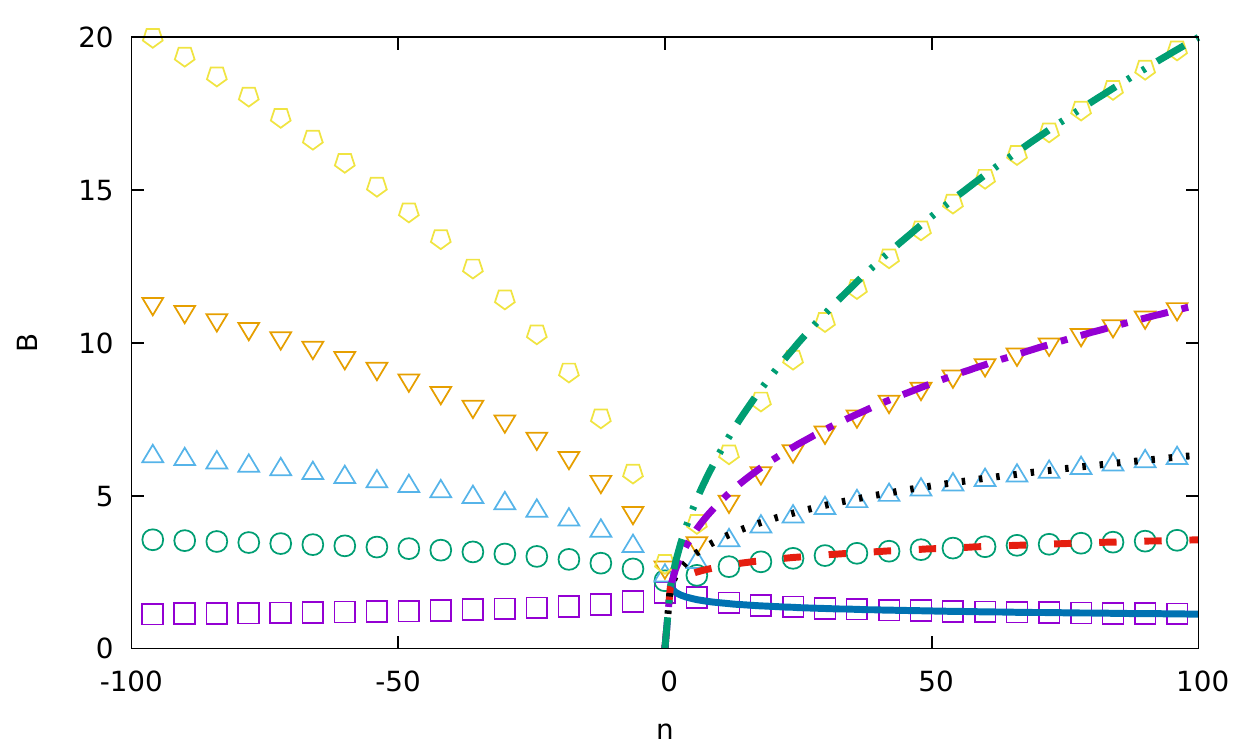}
\caption{Comparison of numerical recursive solutions with smooth ODE solutions 
of the $B$ sequence for $\lambda=1,3,4,5,6$. Lines 
are ODE solutions, dots are numerical solutions.}
\label{Bcomp}
\end{figure}

Such a comparison for both the $A$ and $B$ sequences appears in Figs.~\ref{Acomp} and 
\ref{Bcomp}. Note that the numerical solutions were obtained by recursing backward from 
large values of $k, n$. The starting values for the recursion were chosen from the analytic 
differential equation solutions themselves for the purpose of close agreement in the 
semi-classical limit.  

It is clear that the solutions agree very well in the regimes of large $k, n$, 
but begin to deviate at smaller values. This is expected behavior, of course: 
Quantum corrections are expected to be significant only in the small $k, n$ domain. 

An important observation worth making here is that the $B$ numerical solution 
recurses right through $n = 0$ with no difficulty. Recall that this is the location 
of the classical singularity. This suggests that the classical singularity is 
fully resolved in this model. 
This strongly suggests that in the full 2D model, one is likely to observe that 
the $\tau = 0$ classical singularity is {\em resolved} i.e. the evolution passes 
right through it as it does in the separated solution. 

\section{Full 2D numerical solution}
We now turn our attention to the full two dimensional quantum difference equation in 
the CS model. As done before for the separable solutions, we begin with a stability 
analysis of the CS model. This will allow us to check whether the instability issues we 
uncovered in the separable sequence $A$ manifest themselves in some form in the full 2D 
case as well. 

\subsection{Stability analysis}
We perform a traditional von-Neumann stability analysis of the full 2 model in 
this section. To make the analysis simpler, we make the approximation that $k,n \gg 1$, 
and decompose the solution as:
\begin{equation}
\Psi_{n,k}\rightarrow v^{n}e^{i {(2\pi)\over \lambda} k} ~.
\end{equation}
Substituting this form for the solution and simplifying significantly yields a 
polynomial of the form below for $v$: 
\begin{equation}
v^2 + i\frac{k}{n}[\frac{\sin^2({4\pi\over 
\lambda})+\gamma^2\delta^2_b}{2\sin({4\pi\over \lambda})}]v-1=0 ~.
\end{equation}
The roots of the above equation can be shown to be bounded by unity only if 
$\delta_b\rightarrow0$ and $k < 4n$. It is rather interesting to note this additional 
{\em constraint of $k < 4n$ arising purely from the full quantum dynamics} of the CS model 
of the Schwarzschild interior. This suggests that there is a large portion of the $k, n$ domain 
that is {\em forbidden} due to quantum effects. A similar condition was also found for the 
black hole interior model presented in Ref.~\cite{ab-bh} and was studied in detail in 
Ref.~\cite{bck} (see the Appendix in that paper).  

The restriction $k < 4 n$ is a condition for stability which arises for the macroscopic black holes in the limit when both $k$ and $n$ are 
large. Let us recall from Sec. II that at the horizon $p_b$ vanishes, i.e. $k$ becomes zero. On the other hand, 
at the central singularity both $p_b$ and $p_c$ vanish. That is, both $k$ and $n$ become zero. Therefore the restriction 
$k < 4n$ is not applicable either near the singularity or near the horizon. It does not affect the allowed regions near the horizon and singularity. Rather, it is valid in the interior of a very large black hole when both $k$ and $n$ are large compared to unity.  Using the maximum allowed value of $p_b$ and $p_c$ in classical theory in the Schwarzschild interior, 
we find that maximum values of $k$ and $n$ are: $k_{\mathrm{max}} = n_{\mathrm{max}} = 4 m^2/\gamma \sqrt{\Delta} \lp^2$. Thus even for the maximum value of $k$, it is possible to choose appropriate states which satisfy $k < 4 n$. Though on one hand, the constraint does not seems too restrictive, it nevertheless forbids states which are spread out in $k$ and $n$, such that $k \geq 4 n$. All such states will result in an unstable evolution in the Schwarzschild interior for CS quantization. However, choosing a sharply peaked state which satisfies $k < 4 n$ would yield a stable evolution, as is for example depicted in Fig. \ref{surface}.

It is interesting that if we reverse the role of $k, n$ and perform the 
stability analysis using the flipped decomposition instead:  
\begin{equation}
\Psi_{n,k}\rightarrow v^{k}e^{i {(2\pi)\over \lambda} n} ~.
\end{equation}
Then after some manipulation, the following expression is obtained:
\begin{equation}
 v^4+\alpha v^3 -2(1+2\gamma^2\delta_{b})v^2-\alpha v +1=0
\end{equation}
where,
\begin{equation}
 \alpha = i\frac{8n\sin({4\pi\over \lambda})}{k}. 
\end{equation}
While this is difficult to solve generally, one can numerically check that 
\begin{equation}
 \delta_b\rightarrow0
\end{equation}
leads to stability in this type of evolution as well. This again implies that the mass 
of the black hole must be very large compared to Planck mass. Note that the constraint 
$k < 4n$ is also necessary for stability in this approach.

\subsection{2D Numerical Implementation}
\begin{figure}[htb!]
 \centering
 \includegraphics[width=\columnwidth]{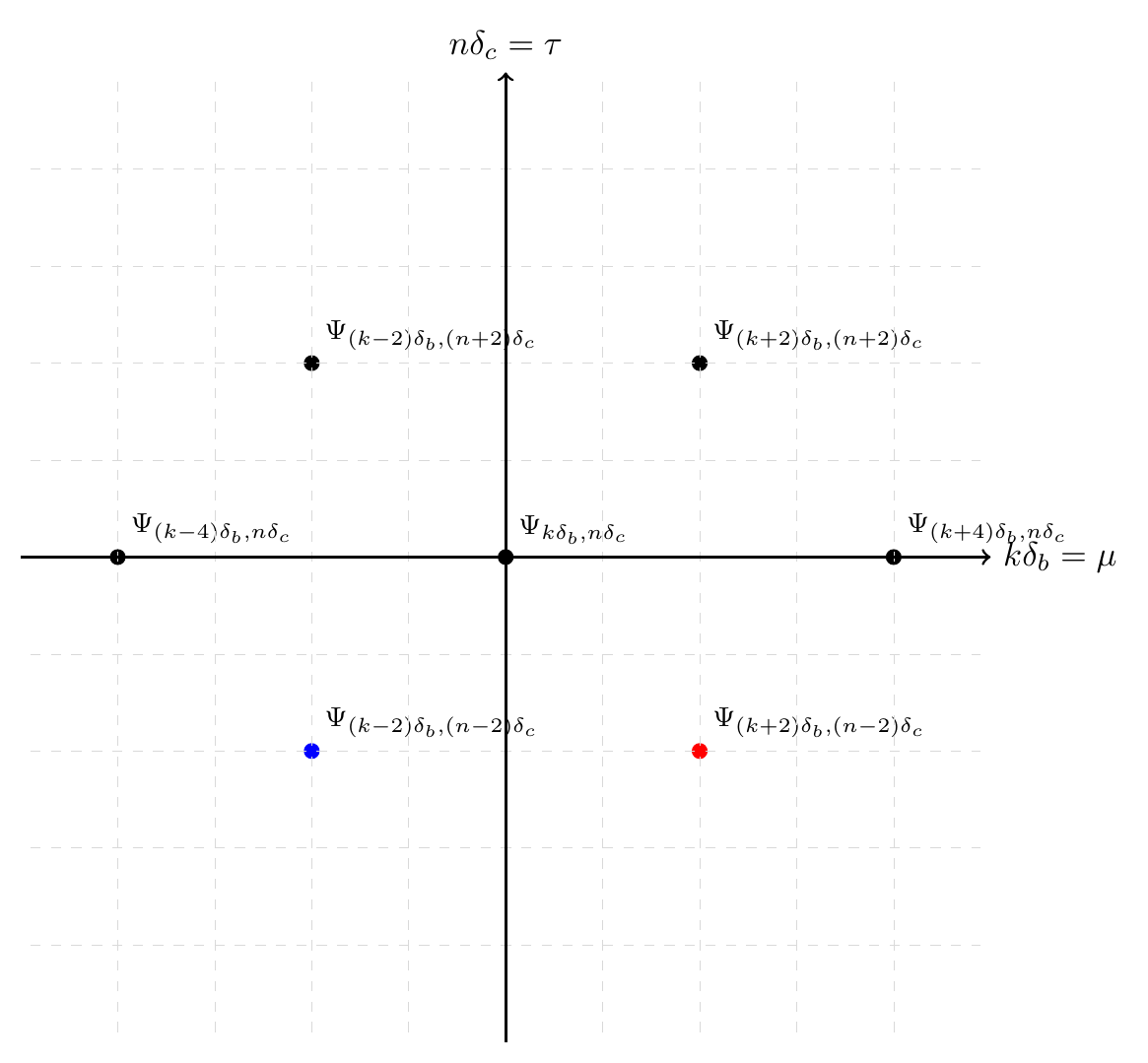}
 \caption{The numerical stencil associated to the 2 model and our evolution 
scheme.}
\label{stencil}
\end{figure}

With the stability analysis completed, we now create a recursive, numerical 
routine to find the solution of the 2 CS model over a wide range of $k, n$. As 
typically done in LQC models, we impose a ``semi-classical'' wave-packet i.e. a 
Gaussian profile for the solution at large values of $k, n$ and then perform a 
backward evolution deep into the quantum regime. To do this, a numerical stencil 
is established to solve for the ``past'', ``back'' values in the positive part of 
the $k$ domain. In the CS model this corresponds to the $\Psi_{k-2,n-2}$ values 
respectively. If we translate the equation of interest into a numerical stencil 
and it takes the form of a seven-point finite differencing scheme. This is 
represented schematically in Fig.~\ref{stencil}.

It is worth commenting on the fact that the solution we obtain should be {\em symmetric} 
i.e. possess mirror symmetry about both $k = 0$ and $n = 0$. The $k = 0$ symmetry can be 
imposed by starting with a double Gaussian wave-packet that possesses the same symmetry and 
using a {\em symmetric} evolution scheme. More specifically, instead of solving for $\Psi_{k-2,n-2}$ 
throughout, the solver routine recurses until it hits $k = 4$. Simultaneously, it also 
begins a similar evolution starting at large negative values of $k$ using $\Psi_{k+2,n-2}$ 
until it reaches $k = -4$. These two phases of the evolution scheme can be executed independently 
and are guaranteed to generate a symmetric result. One can perform a similar procedure for imposing 
the reflection symmetry about $n = 0$ as well. We used all these different approaches and found 
that the final solution we obtain is indistinguishable in these different cases, no matter 
whether or not one imposes these symmetries. 

\subsubsection{Large volume, late time evolution}
To test that this method works, we perform a number of large $k, n$ computations. 
To begin, a Gaussian profile is set up centered in the middle of the positive part of 
the $k$ domain. Next, we evolve this system using the stencil as shown in Fig.~\ref{stencil} 
over the domain of $k\in [0,200]$ and $n \in [100,500]$. The first most basic check 
is to verify  whether the wave-packet maintains its semi-classical Gaussian profile. 
\begin{figure}[htb!]
 \centering
 \includegraphics[width=\columnwidth]{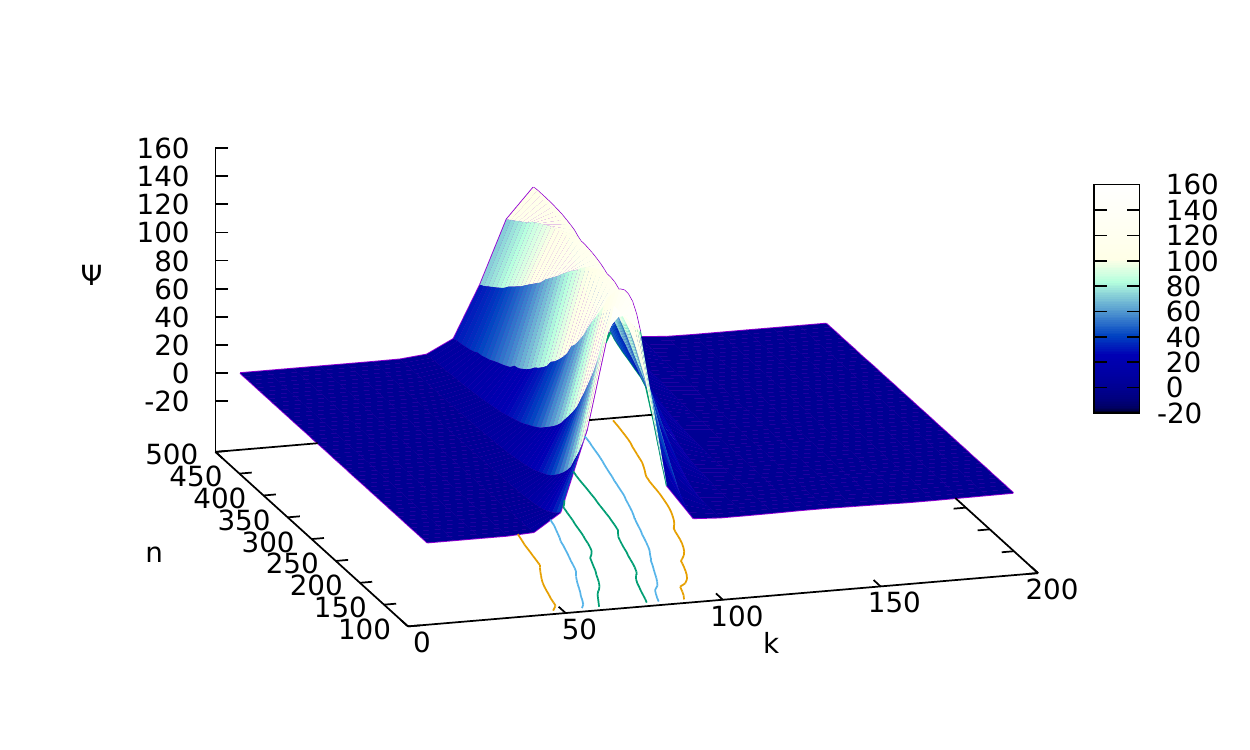}
 \caption{Large volume, late time evolution of a Gaussian wave-packet in 
Schwarzschild interior CS model.}
\label{wavepacket1}
\end{figure}
As can be seen from Fig.~\ref{wavepacket1}, the semi-classical profile remains 
preserved through the evolution.  With the confidence that the Gaussian remains well-preserved 
in its evolution; we can now examine, in detail, the trajectory taken by the Gaussian 
wave-packet. 

\subsubsection{Trajectory of Gaussian wave-packet}
Ref.~\cite{CorSin2016} also presents the semi-classical trajectories in the Schwarzschild 
interior at the late times using the classical Hamiltonian evolution. In the proper 
regime, the trajectory taken by the Gaussian wave-packet in our evolutions must follow the 
same effective equations. To verify this, we first parametrize the trajectory taken by the 
expectation value of $k$ of the Gaussian wave-packet by using a fitting function with a 
form inspired by the expected classical trajectories. To ensure a good fit we ran for a large 
number of iterations in $n,k$.
\begin{figure}[htb!]
 \centering
 \includegraphics[width=\columnwidth]{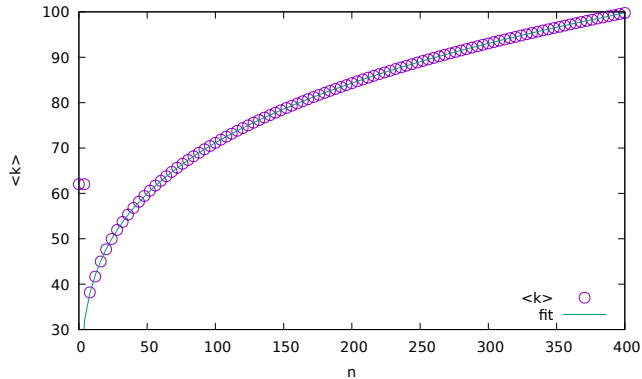}
 \caption{Long time run of trajectory of a Gaussian wave-packet, with classical trajectory fit superimposed.}
\end{figure}
Excellent agreement is observed, except when the evolution is deep in the Planck regime. 
We emphasize that since our computations are performed with $\delta_b \rightarrow 0$, in order 
to compare with Ref.~\cite{CorSin2016}, we must set the mass of the black hole to be very large, 
since $\delta_b$ is inversely related to the horizon radius.

\subsubsection{Singularity resolution}
In order to evolve through the (classical) singularity, we begin by choosing an appropriate 
domain. To reduce numerical errors due to finite-precision computations, we limit the size of 
our computational domain. We ran our computations on a domain of $k,n\in[-500,500]$ and also 
$[-400, 400]$ with a Gaussian wave-packet starting at $k=\pm 100$. Surface plots of 
the entire evolution is depicted in Figs.~\ref{surface}.
\begin{widetext}
\begin{center}
\begin{figure}[htb!]
 \includegraphics[width=0.45\columnwidth]{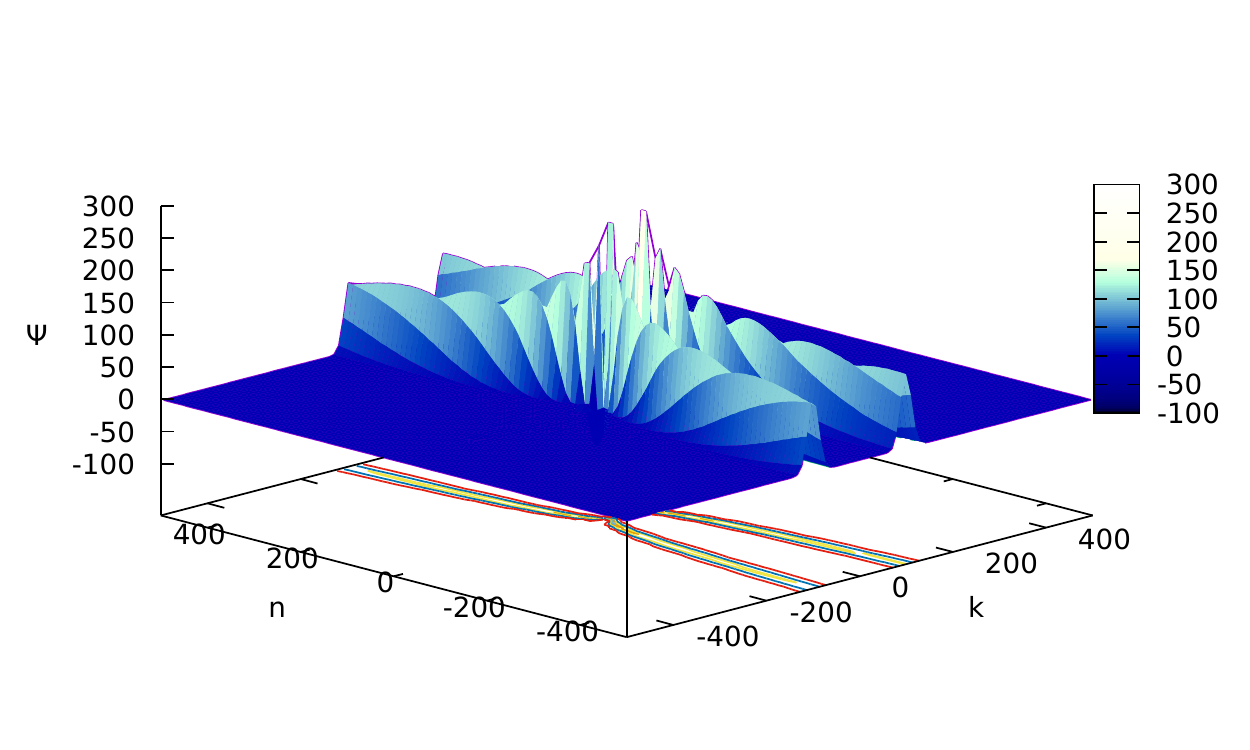}
 \includegraphics[width=0.45\columnwidth]{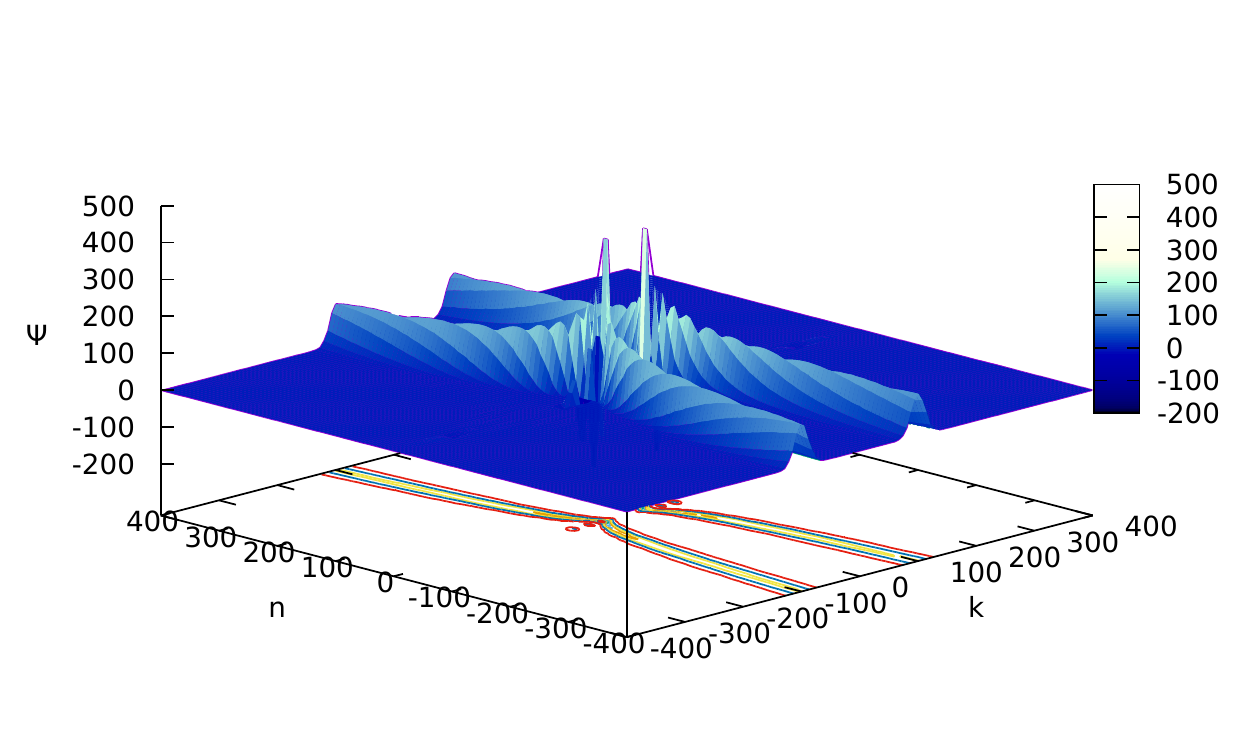}
 \caption{Surface plot of a Gaussian wave-packet evolving through singularity 
in a square $[-500, 500]$ and $[-400, 400]$ domain.}
\label{surface}
\end{figure}
\end{center}
\end{widetext}

\begin{figure}[htb!]
 \centering
 \includegraphics[width=\columnwidth]{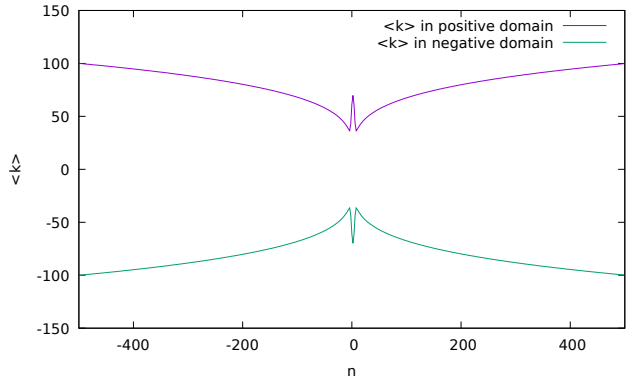}
 \caption{Plot of expectation values of $k$ with a 
``bounce'' observed as it passes through singularity}
\label{kexpec}
\end{figure}

\begin{figure}[htb!]
 \centering
 \includegraphics[width=\columnwidth]{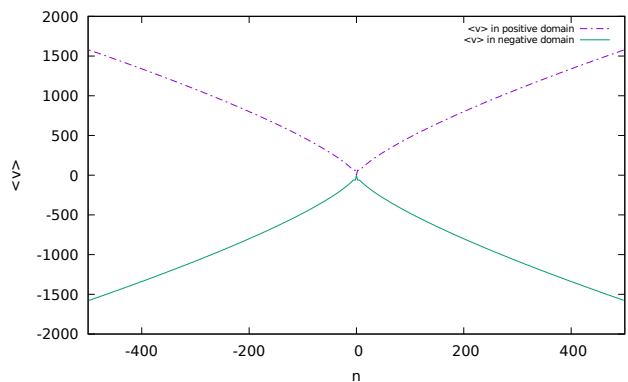}
 \caption{Expectation values of `volume' $v$ given by the ratio of physical volume and a constant factor $2 \pi \gamma^{3/2} \lp^3$ is plotted.}
\label{expecv}
\end{figure}

In addition, in Fig.~\ref{kexpec} we show the expectation values of $k$ for both Gaussian packets 
throughout the entire duration of the evolution. We also plot the expectation value of volume divided 
by a constant factor of $2 \pi \gamma^{3/2} \lp^3$ in Fig. \ref{expecv}. 
It is clear from these figures that the Gaussian wave-packet evolves right 
through the classical singularity at $\tau=0$. Moreover, after passing through the 
singularity, the packet regains its semi-classical trajectory. It is also interesting 
to note the packet appears to ``bounce'' away from $\mu = 0$ and not pass through it. 
\vskip0.3cm

\begin{center}
\begin{figure}[htb!]
 \includegraphics[width=\columnwidth]{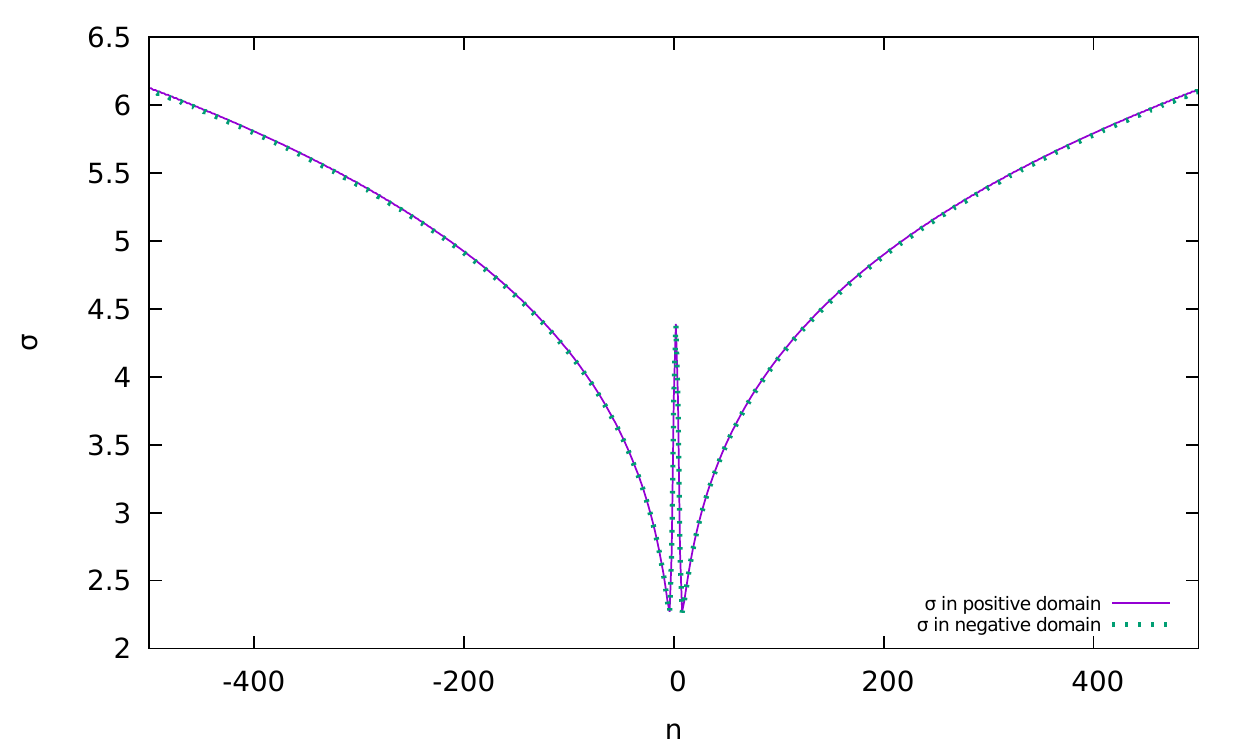}
 \caption{Plot of the width or standard deviation of Gaussian wave-packet over 
the entire evolution. The solid (dashed) curve depicts the width of the Gaussian wave-packet that is localized in the positive (negative) $k$ subdomain.}
\label{stddev}
\end{figure}
\end{center}

A plot of the behavior of the Gaussian wave-packet's width (computed as the 
standard deviation of the distribution) is presented in Fig.~\ref{stddev}. In this figure, the solid curve shows the dispersion for positive values of $k$. The dotted curve shows the dispersion for the negative values of $k$. Their agreement shows that the evolution of spread of the state is symmetric and $k$ and $n$.

It is clear that the width of the wave-packet grows dramatically as it 
approaches the deep quantum regime. This is quite reasonable, given that near 
the classical singularity we expect the solution to have large quantum 
corrections; in fact, we do not expect the solution to even resemble a 
semi-classical wave-packet in that regime! It is, of course, striking that even 
after such a quantum state has developed, evolving the system further 
eventually results in another transition to semi-classical behavior once the 
system emerges on the other side of the deep quantum regime.  

\section{Summary and conclusions}
The goal of our analysis was to understand some of the main properties of the 
quantum Hamiltonian constraint in the loop quantization of the Schwarzschild 
interior as proposed in Ref.~\cite{CorSin2016}. Till date, this and its 
generalization recently studied in Ref.~\cite{oss} are the only known loop 
quantizations of the Schwarzschild interior which are free from dependence on the 
fiducial structure and give correct infra-red limit in agreement with GR. In 
our study we focused on the von-Neumann stability of the quantum constraint -- 
a quantum difference equation in two variables $n$ and $k$ which measure the 
triads $p_b$ and $p_c$, and on the resolution of the central singularity which 
occurs when both $n$ and $k$ vanish. Von-Neumann stability of various loop quantized 
spacetimes has been studied earlier and has provided important insights on the 
viability of the underlying quantization, in particular on the way the quantum 
dynamics approximates the classical solutions at large scales.   In this manuscript, 
von-Neumann stability was analyzed in two different ways. The first method was based on 
separation of variables by expressing the quantum constraint in two quantum 
difference equations independently in $n$ and $k$. The second method used the 
two-dimensional quantum constraint in its original coupled form. Results from 
both the methods turn out to be the same. We found that the quantum Hamiltonian 
constraint is stable only if $\delta_b$, one of the discreteness parameters of 
the quantization, vanishes and $k < 4 n$.  The latter inequality arises purely 
from the stability of the quantum difference equation. A similar inequality has been noted before in previous 
loop quantizations of the black hole interior \cite{CarKha2006,bck}, but has no known 
parallel in the loop quantization of any other types of spacetimes.  The inequality does not apply near the horizon or the central singularity, but is only valid in the interior of a large black hole.  The above inequality implies that 
only those states are compatible with the quantum constraint which are localized 
during the entire evolution such that $k$ is always less than $4 n$. It turns out 
that the maximum allowed values of $k$ and $n$ are equal, and since the quantum difference 
equation couples points with a step difference of two in $k$ and $n$, this is not a 
severe restriction. Indeed, sharply peaked Gaussian states can be successfully 
evolved choosing an appropriate  numerical grid as demonstrated in our 
analysis. This restriction implies that studying evolution with more general 
states will require an extra care, and some types of states which are not 
localized such that $k < 4n$ are forbidden. It is interesting to note that 
such a restriction on the allowed states can be extracted from the stability 
analysis without any prior knowledge of the physical Hilbert space. In future work, it will be interesting to understand whether such a constraint arises independently from the physical Hilbert space.

The constraint on the discreteness parameter $\delta_b$ for a stable evolution is an 
important one.   Without giving up the underlying kinematical properties of the 
quantization, this constraint is true only when the mass of the black hole becomes 
very large compared to the Planck mass. Thus, for the black holes of the astrophysical 
interest, the quantum Hamiltonian constraint provides a stable evolution. It is  
to be noted that though the stability analysis on one hand signals that for black 
holes with masses not large compared to Planck mass instabilities arise, it is not 
clear whether this is a no-go result. The reason is tied to the underlying approximation 
in the stability analysis which requires large values of $k$ and $n$. For a given 
black hole of mass $m$, the maximum allowed values of $k$ and $n$ are proportional to 
$m^2$. Therefore, it is only for the large black hole masses that one can consider large values 
of $k$ and $n$. For such black holes the value of $\delta_b$ is extremely small. In this sense, for the large black hole masses the 
 constraint of the discreteness parameter 
is in a way a consistency condition. It is to be noted that for small black hole masses, the approximation in the 
stability analysis is not strictly valid. Though analytically one can not conclude the instability for small black holes, its existence is confirmed in the numerical simulations. Thus,  
there is evidence of instability 
independent of the von-Neumann analysis. Further work on analytical understanding of the 
properties of the quantum difference equation for small mass black holes is 
needed for a complete picture.

In the limit where the black hole mass is much larger than the Planck mass, we 
investigated the fate of the singularity resolution by using Gaussian initial 
states peaked at the classical trajectories at large values of $n$ and $k$. The 
states when evolved using quantum constraint are found to be peaked on the classical 
trajectory for a long time almost up to near the classical singularity. Our numerical 
simulations show that the quantum evolution never breaks down and the states evolve 
through central singularity. The quantum geometric discreteness results in a resolution 
of the classical central singularity. The singularity resolution is such that the 
magnitude of the expectation value of $k$ takes a non-zero minimum which confirms 
with the picture of bounce, here for the $p_b$ triad. However, there is no bounce in 
the triad $p_c$. Rather the evolution is such that the wavefunction passes through 
the classical singularity at $p_c = 0 $ (or $n =0$). Since the physical volume is 
given by $4 \pi p_b |p_c|^{1/2}$, its expectation values vanish at $n=0$ but the 
evolution continues from positive to negative values of $n$. The magnitude of the 
expectation values of $k$ and volume reveal a symmetric picture across $n=0$. This 
happens because the passage through the singularity does not affect the semi-classical 
properties of the Gaussian state and the classical solution is obtained on two sides 
of the central singularity. This result confirms  the black hole to white hole 
transition in Ref.~\cite{CorSin2016}. 

Let us now discuss some caveats of our analysis. An important limitation is 
that we lack knowledge of the spectrum of the quantum Hamiltonian 
constraint and the physical inner product in this quantization. It should be noted 
that this limitation is not peculiar to the quantization studied here~\cite{CorSin2016}, 
but is shared by all known loop quantizations of the black hole interior~\cite{ab-bh,modesto,bv,gp1,gp2}. 
Further, it does not affect stability analysis or resulting conclusions but can 
potentially modify the details of singularity resolution. This is because  the 
computation of the expectation values of $k$ and volume are computed using the 
kinematical $L^2$ norm and not the physical norm. It is quite possible that 
expectation values of the self-adjoint observables using physical inner product yield 
a picture of singularity resolution which deviates from the one in our analysis and 
is more in agreement with bouncing of both the triads as in the loop quantization of 
the Bianchi-I spacetime \cite{bianchi-madrid,bianchi-lsu}. The Gaussian states used in 
our analysis do not take into account the allowed spectrum of the quantum constraint. 
In a sense our simulations on singularity resolution should thus be seen as preliminary, 
in the same way earlier works in isotropic LQC \cite{coordinate,emergence}, which 
were later improved once the physical Hilbert space became available \cite{aps3}. It 
will be interesting to understand the way the constraint $k < 4 n$ found from the 
stability analysis emerges from the knowledge of the spectrum and the physical inner 
product.

One main result of our analysis is that there are some stability issues with 
the quantum constraint of the loop quantization of Schwarzschild interior 
proposed in Ref.~\cite{CorSin2016}. The stability analysis leads to a constraint that only macroscopic black holes result in a stable evolution. Though this can have various 
implications, we should first 
explore whether such a constraint arises in the physical Hilbert 
space. In LQC, a similar situation arose in the closed isotropic model where early results found 
instabilities of the difference equation \cite{green}. It turned out that this was a result of an incomplete knowledge of the Hilbert space 
and the choice of eigenfunctions which were needed to be chosen very carefully because of the discrete spectrum of the quantum Hamiltonian 
operator \cite{apsv}. Such a discrete spectrum typically arises in LQC when the spatial manifold is bounded, for example in the spatially closed 
model \cite{apsv}. The Schwarzschild interior shares similar features and 
it is possible that the corresponding quantum Hamiltonian operator has a discrete spectrum in the physical Hilbert space. If so, it is possible that 
instabilities are cured once  eigenfunctions are chosen carefully to construct physical states in numerical simulations. If the 
spectrum of the quantum Hamiltonian approaches a continuum for large black holes, then such a spectrum would not effect our results for the 
large mass black holes. In such a case, a potential explanation for an instability only for the small black 
holes can arise. But, to answer these questions reliably, one needs to obtain the physical Hilbert space in loop quantized black hole 
spacetimes. Our results lead to an emphasis on such investigations in future works.  

In case the instabilities for small masses are found 
even at the level of the physical Hilbert space then the viability of CS quantization becomes questionable. 
Some peculiar features of this quantization 
have been noted earlier, which include a large difference in the white hole 
mass from the parent black hole mass after the singularity is resolved \cite{CorSin2016}. A 
generalization of this quantization has been recently proposed which results in 
a symmetric bounce in terms of the masses of the black and white holes by modifying 
the way quantum discreteness enters the quantum difference equation \cite{oss}. As emphasized 
earlier these are the only two viable quantizations of loop quantized Schwarzschild 
interior. Here important questions are whether the quantum difference equation in this 
generalized quantization is von-Neumann stable, and the details of the singularity resolution. It will be interesting to understand whether even this quantization results in a 
similar conclusion about instability for black holes with small masses. If the latter turns out to be generic feature of 
different loop quantizations, then is quantum gravity telling us something fundamental about the existence of small black holes? Future 
investigations 
on these issues promise important insights on these fundamental questions. 



\section*{Acknowledgements} We are grateful to an anonymous referee for valuable suggestions for the improvement of the presentation of our 
manuscript. A.Y. and G.K. acknowledge research support from UMass Dartmouth, NSF Grants No. 
PHY-1414440 and No. PHY-1606333, 
and from the U.S. Air Force agreement No. 10-RI-CRADA-09. P.S. is supported by 
NSF grants  PHY-1404240 and PHY-1454832.


\begin{thebibliography}{10}

\bibitem{mb-livrev}
M.~Bojowald,  {Living Rev. Rel.}, {\bf 11}, 4 (2008)


\bibitem{as-status}
A.~Ashtekar and P.~Singh,  {Class. Quant. Grav.}, {\bf 28}, 213001 (2011).


\bibitem{aps3}
A.~Ashtekar, T.~Pawlowski, and P.~Singh, {Phys. Rev.} D{\bf 74}, 084003 (2006). 
 
  
  
\bibitem{wide}   P.~Diener, B.~Gupt, M.~Megevand, and P.~Singh,  Class.
  Quant. Grav., {\bf 31}, 165006 (2014).
  
\bibitem{slqc} A.~Ashtekar, A.~Corichi and P.~Singh,
  Phys.\ Rev.\ D {\bf 77}, 024046 (2008)
  
\bibitem{craig}   D.~A.~Craig,
  Class.\ Quant.\ Grav.\  {\bf 30}, 035010 (2013)

\bibitem{BriCarKha2012} D. Brizuela, D. Cartin, G. Khanna, SIGMA {\bf 8}, 001 
(2012).

\bibitem{Sin2012} P. Singh, Class. Quantum Grav. {\bf 29}, 244002 (2012).  




\bibitem{bianchi-madrid} M.~Martin-Benito, G.~A.~M.~Marugan and T.~Pawlowski,
  Phys.\ Rev.\ D {\bf 80}, 084038 (2009)


\bibitem{bianchi-lsu} P.~Diener, A.~Joe, M.~Megevand and P.~Singh,
  Class.\ Quant.\ Grav.\  {\bf 34}, 094004 (2017)

\bibitem{generic} P.~Singh,
  Class.\ Quant.\ Grav.\  {\bf 26}, 125005 (2009); P.~Singh,
  Phys.\ Rev.\ D {\bf 85}, 104011 (2012); S.~Saini and P.~Singh,
  Class.\ Quant.\ Grav.\  {\bf 33}, no. 24, 245019 (2016); S.~Saini and P.~Singh,
  Class.\ Quant.\ Grav.\  {\bf 34}, no. 23, 235006 (2017)

  
\bibitem{ab-bh} A.~Ashtekar and M.~Bojowald,  Class. Quant. Grav. {\bf 23}, 391 
(2006).

\bibitem{modesto}
L.~Modesto,  Class. Quant. Grav. {\bf 23}, 5587 (2006).


\bibitem{CarKha2006} D. Cartin, G. Khanna, Phys. Rev. D {\bf 73}, 104009 (2006).


\bibitem{bv} C.~G.~Boehmer and K.~Vandersloot,  Phys.\ Rev.\ D {\bf 76}, 104030 
(2007)

\bibitem{bck} M.~Bojowald, D.~Cartin and G.~Khanna,
  Phys.\ Rev.\ D {\bf 76}, 064018 (2007)


\bibitem{gp1} M.~Campiglia, R.~Gambini, and J.~Pullin,  {AIP Conf. Proc.},
  {\bf 977}, 52 (2008).
  


\bibitem{SabKha2007} S. Sabharwal, G. Khanna, Class. Quantum Grav. {\bf 25}, 
085009 (2008).

\bibitem{gp2} R.~Gambini and J.~Pullin,  {\em Phys. Rev. Lett.} {\bf 110}, 
211301 (2013).


\bibitem{CorSin2016} A. Corichi, P. Singh, Class. Quantum Grav. {\bf 33}, 
055006 (2016).


 \bibitem{prob} A.~Corichi and P.~Singh,
  Phys.\ Rev.\ D {\bf 78}, 024034 (2008); W.~Nelson and M.~Sakellariadou,
  Phys.\ Rev.\ D {\bf 78}, 024006 (2008) 
  
\bibitem{preclassicality} M. Bojowald, Phys. Rev. Lett. {\bf 87}, 121301 
(2001); M. Bojowald and G. Date, Class. Quant. Grav. {\bf 21}, 121 (2004).   
  
  
\bibitem{chimera}   P.~Diener, B.~Gupt and P.~Singh,
  Class.\ Quant.\ Grav.\  {\bf 31}, 025013 (2014)

  




\bibitem{coordinate} M.~Bojowald, P.~Singh and A.~Skirzewski, Phys.\ Rev.\ D 
{\bf 70}, 124022 (2004)

\bibitem{emergence} P.~Singh and K.~Vandersloot,
  Phys.\ Rev.\ D {\bf 72}, 084004 (2005).
    
\bibitem{aag}   I. Agullo, A. Ashtekar, B. Gupt, Class. Quant. Grav. {\bf 34}, 074003 (2017) 
  
\bibitem{cyclic} P. Diener, B. Gupt, M. Megevand, P. Singh, ``Numerical simulations with negative potentials in loop quantum cosmology,'' 
To appear (2017).  



\bibitem{green} D. Green, W.G. Unruh, Phys.Rev. D {\bf 70},  103502 (2004).

\bibitem{apsv} A. Ashtekar, T. Pawlowski, P. Singh, K. Vandersloot,  Phys. Rev. D {\bf 75}, 024035 (2007). 


  
\bibitem{CarKhaBoj2004} D. Cartin, G. Khanna, M. Bojowald, Class. Quantum Grav. 
{\bf 21}, 4495 (2004).






\bibitem{oss} J.~Olmedo, S.~Saini and P.~Singh, Class. Quantum Grav. {\bf 34} 225011 (2017).









\end{thebibliography}
\end{document}